\documentclass[prd,reprint,amsmath,amssymb,floatfix,aps,longbibliography,superscriptaddress]{revtex4-2}
\usepackage{graphicx}
\usepackage{dcolumn}
\usepackage{bm}
\usepackage[colorlinks=true,urlcolor=blue,anchorcolor=black,citecolor=blue,filecolor=black,linkcolor=blue,menucolor=blue,linktocpage=true]{hyperref} 
\usepackage{subfigure}
\usepackage{natbib}

\usepackage{subfigure}

\usepackage{booktabs}
\usepackage{colortbl}
\usepackage{collcell}
\usepackage{enumerate}
\usepackage{enumitem}
\usepackage{float}
\usepackage{verbatim}
\usepackage{multirow}
\usepackage{xparse}
\usepackage{tabularx}
\usepackage{cancel}

\usepackage{multirow}

\newcolumntype{K}[1]{>{\centering\arraybackslash}m{#1}}

\newcommand{\Mr}{\mathbf{M}_{\rm R}}
\newcommand{\Yl}{\mathbf{Y}^\ell}
\newcommand{\Ynu}{\mathbf{Y}^\nu}
\newcommand{\Mnu}{\mathbf{M}^\nu}
\newcommand{\Ur}{\mathbf{U}_{\rm R}}

\newcommand{\eps}{\boldsymbol\epsilon}

\usepackage[normalem]{ulem}
\usepackage{soul}
\setstcolor{red}

\definecolor{rossos}{cmyk}{0,1,1,0.55}
\definecolor{mygreen}{rgb}{0.27, 0.64, 0.48}
\definecolor{mygray}{gray}{0.95}

\begin{document}

\title{Large CP Violation from the Minimum Seesaw Model}

\author{Yu-Cheng Qiu}
\email{ethanqiu@sjtu.edu.cn}
\affiliation{Tsung-Dao Lee Institute, Shanghai Jiao Tong University, Shanghai, 201210, China}

\author{Jin-Wei Wang}
\email{jinwei.wang@uestc.edu.cn}
\affiliation{School of Physics, University of Electronic Science and Technology of China, Chengdu 611731, China}

\author{Tsutomu T. Yanagida}
\email{tsutomu.tyanagida@sjtu.edu.cn}
\affiliation{Tsung-Dao Lee Institute and School of Physics and Astronomy, \\
Shanghai Jiao Tong University, 520 Shengrong Road, Shanghai, 201210, China}
\affiliation{Kavli IPMU (WPI), The University of Tokyo, Kashiwa, Chiba 277-8583, Japan}

\begin{abstract}
The minimum seesaw model with two right-handed neutrinos is considered, where the lightest neutrino is naturally massless. Instead of adopting texture zeros in the lepton Yukawa matrices, which cause both theoretical and experimental troubles, here we propose two-$\eps$ textures, where $\eps$ is a small number. Combined with neutrino oscillation experimental data, we find that a large CP angle is preferred for the normal neutrino mass order. In contrast, the CP angle almost vanishes for the inverted order. This can be well tested in near-future experiments, such as Hyper-Kamiokande. Besides, the predicted effective Majorana neutrino mass $m_{ee}$ and the total neutrino mass $\sum m^\nu_i$ are also within reach of ongoing or future experiments.
\end{abstract}

\maketitle

\section{Introduction}
The super heavy Majorana right-handed (RH) neutrinos $N_i$ are very attractive since they explain naturally the observed small masses of the neutrinos via the seesaw mechanism~\cite{Minkowski:1977sc,Yanagida:1979as,Yanagida:1979gs,Gell-Mann:1979vob} and the observed baryon-number asymmetry in the universe by the decays of the super heavy neutrinos $N_i$ (dubbed the leptogenesis)~\cite{Fukugita:1986hr,Buchmuller:2005eh}. It was pointed out a long time ago that the presence of two heavy Majorana neutrinos $N_{1,2}$ is enough to explain all the above observations~\cite{Frampton:2002qc,Endoh:2002wm}. We call it the minimum seesaw model. A crucial prediction of this model is the vanishing mass of the lightest neutrino and there is no cancellation of the neutrinoless double-beta decay amplitudes, 
e.g., $m_{ee}\sim 1.3$--$3.9$ meV for Normal Order (NO) and $m_{ee}\sim 17$--$49$ meV for Inverted Order (IO)~\cite{ParticleDataGroup:2022pth}.

Based on the minimal seesaw model, one of us (T.T.Y.) further assumed two zeros in the Yukawa coupling matrix of the heavy neutrinos to expect further predictions testable at low energies based on Occam's razor approach\footnote{The requirement for the successful leptogenesis demands at least two nonzero elements. However, for the two or three nonzero cases, two vanishing neutrino mixing angles are predicted, which is contradicted by the observations. Therefore, at most two zero elements are allowed.} \cite{Harigaya:2012bw}. They predicted the inverted mass hierarchy in the neutrino mass matrix and the maximal CP violating phase ($\delta_{\rm CP} \simeq \pm \pi/2$) in the neutrino oscillation, which is consistent with the results of other detailed follow-up studies \cite{Barreiros:2018ndn}. However, there are some unsatisfactory theoretical and experimental aspects to these results. 
From an experimental point of view, the global fit analyses of current neutrino oscillation experiment data provide a $2$--$2.7~\sigma$ preference
for the normal ordering~\cite{deSalas:2020pgw,Esteban:2020cvm,Capozzi:2021fjo}, which contradicts with the above results. From a theoretical point of view, the exact two texture zeros in Ref.~\cite{Harigaya:2012bw} are not natural. Specifically, these two zero elements in the Yukawa matrix should be located in different columns and different rows~\cite{Harigaya:2012bw,Barreiros:2018ndn}, which is not easy to achieve within a natural fundamental theory.

For the above reasons, we propose to relax the exact zeros by replacing them with small parameters $\eps$  
and see if the above issues can be avoided and if there are other testable predictions \footnote{Note that this $\boldsymbol\epsilon$ texture can be easily achieved within the Froggatt-Nielsen (FN) model. Interested readers can refer to our previous paper \cite{Qiu:2023igq}, where we give a detailed example.}. 
This attempt is intriguing, because (1) the nonzero $\eps$ are no longer constrained by any position; (2) the small value of $\eps$ may introduce a new hierarchy for the Yukawa matrix, which is common within the Standard Model framework.  
Interestingly, by adopting the two-$\eps$ texture of the Yukawa matrix in the minimum seesaw mechanism, we find that the Normal Order case naturally has a large CP violation ($\sim \pm \pi/2$), while the Inverted Order case has a small one ($\sim 0$ or $2\pi$).  These predictions can be tested by near-future experiments, such as Hyper-Kamiokande~\cite{HyperKamiokande:2018ofw}.\\


\section{Minimum Seesaw Model}\label{sec:msm}
We are considering the minimum seesaw model, which contains only two heavy RH neutrinos. Note that two heavy RH neutrinos are enough for the leptogenesis to work.
Considering only Yukawa and mass terms, the lepton Lagrangian density with RH neutrino fields is
\begin{equation}
\mathcal{L}=\overline{\ell_{\rm L}}\Yl \Phi\,e_{\rm R} + \overline{\ell_{\rm L}}\Ynu \tilde{\Phi}\nu_{\rm R} -\frac{1}{2}\overline{(\nu_{\rm R})^c}\Mr\nu_{\rm R} + \text{H.c.}\,,
\end{equation}
where $\ell_{\rm L}^j$ and $e_{\rm R}^j$ are 
$SU(2)_{\rm L}$ lepton doublets and singlets,
$\Phi$ is the Higgs doublets,
$\tilde{\Phi}=i\sigma_2 \Phi^\ast$ is its dual.
$\nu_{\rm R}^k$ is heavy RH neutrinos, SM singlets.
Here $j=1,2,3$ and $k=1,2$ are the generation labels, whose summation should be understood in the $\mathcal{L}$.
The $\Ynu$ and $\Mr$ represent the Dirac neutrino Yukawa couplings and RH neutrino Majorana mass matrices. Hereafter, we take a basis where the $\Mr=\text{diag}(M_1, M_2)$ and $\Yl$ are diagonal. Without loss of generality, we take $M_1$ and $M_2$ as real and positive numbers, and we arrange $M_1<M_2$.

There are only two RH neutrinos, 
which means that $\mathbf{Y}^{\nu}$ is a $3\times 2$ complex matrix.
After integrating out the heavy $\nu_R$ degree of freedoms (the seesaw mechanism), the effective Majorana neutrino mass matrix $\Mnu$ can be derived through the seesaw formula~\cite{Minkowski:1977sc,Yanagida:1979as,Gell-Mann:1979vob}
\begin{align}
\Mnu=-v^2\Ynu\Mr^{-1}{\Ynu}^T\,,
\label{eq:Mnuseesaw}
\end{align}
which is valid for $\Mr\gg v$, where $v=174\,{\rm GeV}$ is the vacuum expectation value of the neutral component of $\Phi$. The $\Mnu$ is a symmetric matrix, which can be diagonalized by 
a unitary matrix $\mathbf{U_\nu}$ as
\begin{equation}
    \mathbf{U}_{\nu}^\dagger \Mnu \mathbf{U}_{\nu}^*=\text{diag}(m_1,m_2,m_3) \equiv {\bf d}_{\rm \nu}\,.
\label{eq:Mnudiag}
\end{equation}
One important implication of the minimum seesaw mechanism is that the lightest neutrino mass is zero~\cite{King:1999mb,Frampton:2002qc}, since the rank of $\Mnu$ in Eq.~\eqref{eq:Mnuseesaw} is 2. Thus, the model predicts $m_1=0$ for the normal neutrino mass order and $m_3=0$ for the inverted neutrino mass order~\cite{Harigaya:2012bw}. With the measured neutrino mass square difference from oscillation experiments, one could express three neutrino masses as

\begin{equation}
    \mathbf{d}_\nu^{\rm (NO)} ={\rm diag} \left(0,\sqrt{\Delta m_{21}^2},\sqrt{\Delta m_{3l}^2}\right),
\end{equation}
\begin{equation}
    \mathbf{d}_\nu^{\rm (IO)} = {\rm diag} \left(\sqrt{|\Delta m_{3l}^2|-\Delta m_{21}^2},\sqrt{|\Delta m_{3l}^2|},0\right)\;
\end{equation}
where we have adopted the convention in Ref.~\cite{Esteban:2020cvm}. The values of $\Delta m_{21}^2$ and $|\Delta m_{3l}^2|$ are given in Table~\ref{tab:datatable}. Here $\Delta m_{3l}^2 = \Delta m_{31}^2>0$ for the NO and $\Delta m_{3l}^2 = \Delta m_{32}^2<0$ for the IO. 
Throughout this work, we will use the standard parametrization of the Pontecorvo-Maki-Nakagawa-Sakata (PMNS) matrix (note that we work on diagonal charged lepton basis) ~\cite{ParticleDataGroup:2016lqr},
\begin{widetext}
\begin{gather}
\mathbf{U}_{\rm P}=\mathbf{U}_{\nu}^*=\begin{pmatrix}
c_{12}c_{13}&s_{12}c_{13}&s_{13}e^{-i\delta}\\
-s_{12}c_{23}-c_{12}s_{23}s_{13}e^{i\delta}&c_{12}c_{23}-s_{12}s_{23}s_{13}e^{i\delta}&s_{23}c_{13}\\
s_{12}s_{23}-c_{12}c_{23}s_{13}e^{i\delta}&-c_{12}s_{23}-s_{12}c_{23}s_{13}e^{i\delta}&c_{23}c_{13}
\end{pmatrix}\!\!\begin{pmatrix}
1&0&0\\
0&e^{i\alpha/2}&0\\
0&0&1\\
\end{pmatrix}\,.
\label{Uparam}
\end{gather}
\end{widetext}
Here $s_{ij}=\sin\theta_{ij}$ and $c_{ij}=\cos\theta_{ij}$. The phases $\delta$ and $\alpha$ are Dirac and Majorana CP-violating phases. 
Note that we only have one Majorana phase here
since either $m_1$ or $m_3$ is vanishing in the minimum seesaw model.

One convenient way of parameterizing $\Ynu$ relies on the so-called Casas-Ibarra parametrization~\cite{Casas:2001sr}. By combining Eq.~\eqref{eq:Mnuseesaw} and~\eqref{eq:Mnudiag}, in the basis where both $\Mr$ and $\Yl$ are diagonal, one could write down
\begin{gather}
\mathbf{Y}^\nu=v^{-1}{\bf U}_{\rm P}^*\,\mathbf{d}_{\nu}^{1/2}\,\mathbf{R}\,\mathbf{M}_{\rm R}^{1/2}\;.
\label{eq:CasasandIbarra}
\end{gather}
%
The matrix $\mathbf{R}$ here is a general $3\times 2$ complex matrix,
satisfying $\mathbf{R}\mathbf{R}^T=1$,
which can be parametrized by a single complex angle $z$ in the following way
\begin{gather}
\mathbf{R}^{\text{(NH)}}=\begin{pmatrix}
0&0\\
\cos z&-\sin z\\
\xi \sin z&\xi \cos z 
\end{pmatrix}, \;
\mathbf{R}^{\text{(IH)}}=\begin{pmatrix}
\cos z &-\sin z \\
\xi \sin z &\xi \cos z\\
0&0
\end{pmatrix}\; ,
\label{RmatrixIO}
\end{gather}
with $\xi=\pm 1$. Notice that, in the case of a nondiagonal $\Mr$, the right-hand side of Eq.~\eqref{eq:CasasandIbarra} must be multiplied on the right by $\Ur^\dag$, being $\Ur$ the unitary matrix that diagonalizes $\Mr$.

Since we are investigating the textures in $\mathbf{Y}^{\nu}$, only relative magnitude matters, i.e. $r_{\rm M} = M_1/M_2$. In this paper, we set $M_2 = 10^{14}$ GeV as a benchmark value. Note that the choice of $M_2$ is just an overall factor and will not affect the pattern of $\mathbf{Y}^{\nu}$. 
From Eq.~\eqref{eq:CasasandIbarra}, we could conclude that the absolute value of $3\times2$ Yukawa matrix $|\mathbf{Y}^{\nu}|$ could be determined as a function of a set of parameters $\mathcal{P} = \left\{\theta_{12},\theta_{23},\theta_{13},\Delta m_{21}^2, \left|\Delta m_{3l}^2\right|,\delta,\alpha,r_{\rm M},z, \xi \right\}$.

From Occam's razor approach, Refs.~\cite{Harigaya:2012bw,Barreiros:2018ndn} investigate the patterns in $\mathbf{Y}^{\nu}$ with maximally two zeros. The positions of two zeros in this $3\times2$ matrix are important. 
For example, if two zeros are aligned in the same row or column, the predictions on the neutrino mixing angles would be inconsistent with the observations \cite{Harigaya:2012bw}. Besides, demanding something being zero is very strict since zero could cause serious fine-tuning problems. Here we release these two zeros into small parameters $\eps$s. Once two $\eps$s are in the texture, the positions are no longer matter. For example, we could have them in the same row or column of $\mathbf{Y}^{\nu}$ and still be consistent with observations.

The two-$\eps$ texture of $\mathbf{Y}^{\nu}$ that we are interested in here is equivalent to that there are two elements that are much smaller than others. Assuming that $Y_1$, $Y_2$, and $Y_3$ are the three elements in $\mathbf{Y}^{\nu}$ with the smallest absolute values, and have $|Y_1|\leq |Y_2|\leq |Y_3|$. In this case, the two-$\eps$ texture can be expressed as 
\begin{equation}
    |Y_2|/|Y_3|<\eps,
    \label{eq:criteria}
\end{equation}
where $\eps$ is a small number to make $Y_{1,2}$ distinct with others. We shall show that this criteria is sufficient for selecting out two-$\eps$ texture later.\\

\begin{table}[t!]
\centering
\caption{Neutrino oscillation parameters obtained from the global analysis of Ref.~\cite{Esteban:2020cvm} with SK atmospheric data. Note that $\Delta m_{3l}^2 = \Delta m_{31}^2>0$ for NO and $\Delta m_{3l}^2 = \Delta m_{32}^2<0$ for IO. The $1\sigma$ uncertainties are indicated in super- and subscripts.}
\begin{ruledtabular}
\begin{tabular}{l c c}
\textbf{Parameter} & Normal Order (NO) & ~Inverted Order (IO) \\
\midrule
$\theta_{12}\;(^{\circ})$ & $33.44_{-0.74}^{+0.77}$ & $33.45_{-0.75}^{+0.78}$\\[0.5em]
$\theta_{23}\;(^{\circ})$  & $49.2_{-1.2}^{+0.9}$ & $49.3_{-1.1}^{+0.9}$\\[0.5em]
$\theta_{13}\;(^{\circ})$ & $8.57_{-0.12}^{+0.12}$ & $8.60_{-0.12}^{+0.12}$\\[0.5em]
$\Delta m_{21}^2\;(\text{meV}^2)$ & $74.2_{-2.0}^{+2.1}$ & $74.2_{-2.0}^{+2.1}$\\[0.15cm]
$\left|\Delta m_{3l}^2\right|\;(\text{meV}^2)$ & $2517_{-28}^{+26}$ & $2498_{-28}^{+28}$\\
	\end{tabular}
 \end{ruledtabular}
\label{tab:datatable}
\end{table} 

\section{Monte Carlo Simulation} \label{sec:simulation}
The simulation strategy is quite straightforward. Given a specific $\mathcal{P}$, the $\mathbf{Y}^{\nu}$ can be fixed, by adopting the two-$\eps$ texture criteria~\eqref{eq:criteria}, the sets of parameters $\mathcal{P}$ that meet the requirements can be picked out, then the probability distribution functions (PDF) of CP angles, e.g. $\delta$ and $\alpha$, can be achieved. As shown in Table~\ref{tab:datatable}, the values of mixing angles and neutrino mass square differences are well measured through neutrino oscillation experiments. 
Whereas, the Dirac phase $\delta$ can only be globally fitted, which has a large uncertainty, and the Majorana phase $\alpha$ could not be determined by neutrino oscillation. 
Besides, there are additional three free parameters $r_{\rm M}$, $z$, and $\xi$. The sampling strategy for these parameters is as follows.
\begin{enumerate}[label=(\roman*)]
    \item For the well-measured parameters, namely $\Delta m_{21}^2$, $\left|\Delta m_{3l}^2\right|$, and $\theta_{ij}$, we take their prior distribution as uniform distributions spanning their $1\sigma$ range~\cite{Esteban:2020cvm}.
    
    \item For the unmeasured parameters $\{\delta,\alpha,r_{\rm M},z\}$, we adopt uniform distribution as priors with $\delta \in [-\pi,\pi)$, $\alpha \in [0,2\pi)$, $r_{\rm M} \in [0.01,1]$, $|z|\in [0,10]$ and ${\rm Arg}(z) \in [0,2\pi)$.
    
    \item The $\xi$ appears in Eq.~\eqref{eq:CasasandIbarra} is just a binary parameter, we will randomly sample $\pm1$ with equal probability.
\end{enumerate}
By adopting the above sampling strategy, for each round, we could get one specific $\mathbf{Y}^{\nu}$. 

In Refs.~\cite{Harigaya:2012bw,Barreiros:2018ndn}, due to the help of two-zero texture, the observable quantities $\theta_{13}$ and $\Delta m_{21}^2/\left|\Delta m_{3l}^2\right|$ become predictable. By comparing with the experimental data, they found that the NO scenario is excluded and only IO is valid with large CP angle $\delta = \pm \pi/2$~\cite{Harigaya:2012bw}. However, the two-$\eps$ texture loses this advantage. Applying Eq.~\eqref{eq:criteria}, the desired $\mathbf{Y}^{\nu}$ with two-$\eps$ pattern will be selected. As a result, we could get the distributions of $\{\delta,\alpha,r_{\rm M},z\}$. 

\begin{figure}[t!]
    \centering
    \includegraphics[width=8.6cm]{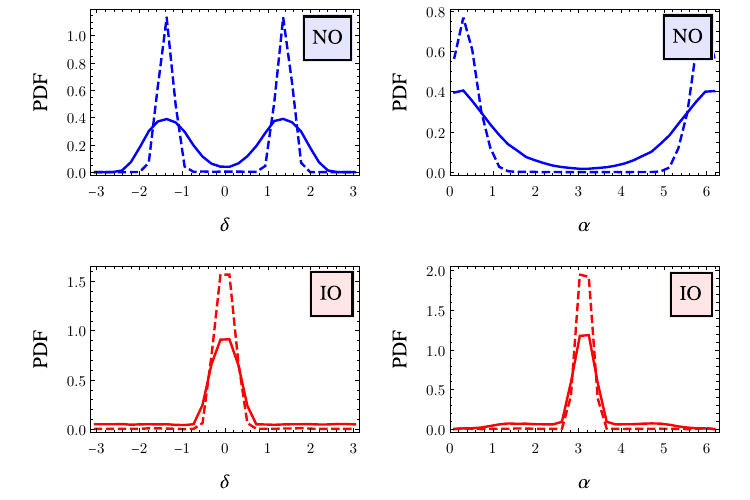}
    \caption{Distributions of $\delta$ and $\alpha$  for both NO and IO cases under two-$\eps$ texture. The solid lines and dashed lines represent $\eps=0.15$ and $\eps=0.11$, respectively.
    }
    \label{fig:delta_alpha}
\end{figure}

In Fig.~\ref{fig:delta_alpha}, we show the PDF of $\delta$ and $\alpha$ for both NO and IO cases. We find that for the NO case the PDF of $\delta$  peaks at the maximum-CP value, i.e., $\delta = \pm \pi/2$, while the Majorana phase $\alpha$ tends to vanish, i.e., $\alpha = 0$ or $2\pi$.  However, for the IO case, the situation is quite different, both the Dirac phase and the Majorana phase tend to vanish, and the peak of PDF of $\delta$ and $\alpha$ are $0$ and $\pi$, respectively. Furthermore, we also demonstrate the effects of different choices of $\eps$, the solid and dashed lines correspond to $\eps=0.15$ and $\eps=0.11$, respectively. It clearly shows that as $\eps$ decreases, the distribution becomes more and more concentrated on peak value.
This indicates that under the constraint of the two-$\eps$ texture of $\mathbf{Y}^\nu$, the NO prefers a large CP angle, while the IO renders a vanishing CP. 
\begin{figure}
    \centering
    \includegraphics[width=7cm]{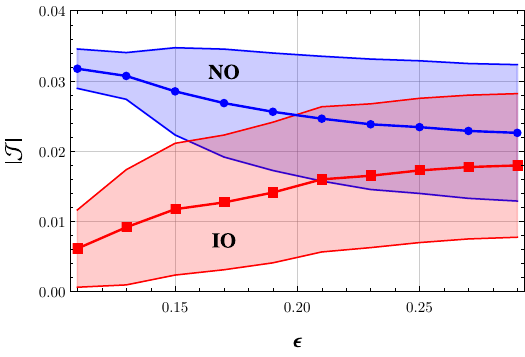}
    \caption{
    The statistics of the Jarlskog $|\mathcal{J}|$, under the two-$\eps$ texture, is plotted against the criteria $\eps$.
    The blue color is for the NO case and the red is for the IO case. Solid lines mark average values and the color bands indicate one standard deviation.}
    \label{fig:J}
\end{figure}
It is known that the size of CP violation can be quantified by the Jarlskog invariant~\cite{Jarlskog:1985ht},
\begin{equation}
    \mathcal{J} = c_{12} s_{12} c_{23} s_{23} c_{13}^2 s_{13} \sin \delta \;,
\end{equation}
where only the Dirac phase is involved.
As shown in Fig.~\ref{fig:delta_alpha}, distribution for $\delta$ is almost $Z_2$-symmetric. 
In Fig.~\ref{fig:J}, we show the absolute value of Jarlskog, $|\mathcal{J}|$ as the function of $\eps$. The blue and red colors correspond to NO and IO cases, respectively. The solid lines represent the average value, while the colored band represents one standard deviation.
One could see that for large $\eps$, the average value of $|\mathcal{J}|$ and the $1\sigma$ band of NO and IO cases almost coincide, which means that there is no difference in CP-violation between NO and IO for the `weak' texture for $\mathbf{Y}^\nu$.
Meanwhile, as $\eps$ goes smaller, the difference between NO and IO becomes larger, and only the NO case possesses a large CP, while the IO tends to have a vanishing CP violation. Another feature is that as $\eps$ gets bigger, the colored band gets wider and wider, which means the distributions become flatter and flatter (see Fig. \ref{fig:delta_alpha}). One can imagine that in the limit of a large $\eps$, the PDF of $\delta$ and $\alpha$ become totally flat and have the largest uncertainties. 
This is quite different from the two-zero texture in Refs. \cite{Harigaya:2012bw,Barreiros:2018ndn}, where only the IO case is consistent with experimental data and with a large Dirac CP angle. This indicates that there is a discontinuity for the PDF of $\delta$ at $\eps=0$. 
This unique feature provides good testability to our model.
The Jiangmen Underground Neutrino Observatory (JUNO) experiment could distinguish NO and IO at $3\sigma$ in the near future~\cite{Stock:2024tmd}. Besides, Hyper-Kamiokande is expected to confirm at the $5\sigma$ confidence level or better if CP symmetry is violated in the neutrino oscillations for $57\%$ of possible $\delta$ values~\cite{HyperKamiokande:2018ofw}. Therefore, the two-$\eps$ texture of the minimum seesaw model could be tested very soon.

\begin{figure}
    \centering
    \includegraphics[width=8.6cm]{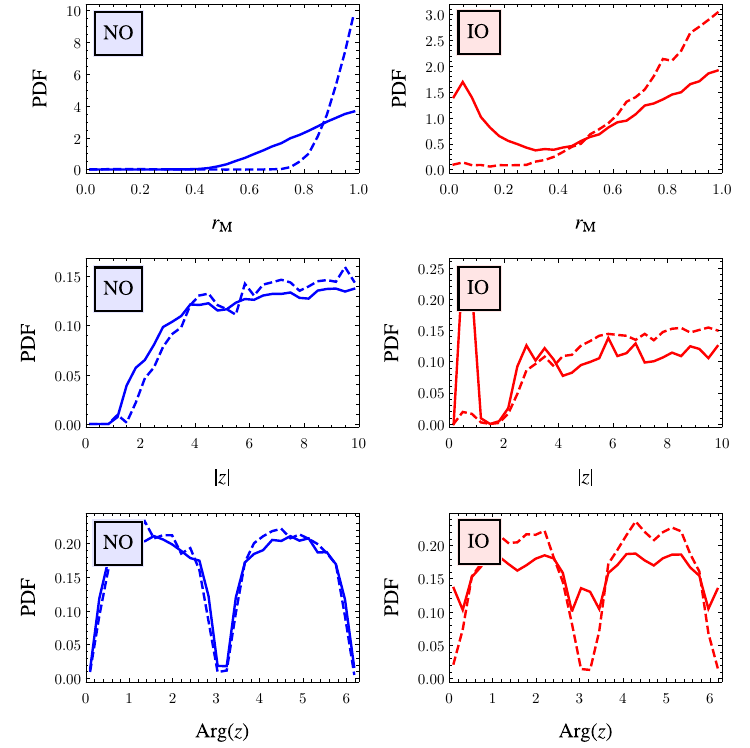}
    \caption{Distributions of $r_{\rm M}$ and $z$  for both NO and IO cases under two-$\eps$ texture. The solid and dashed lines represent $\eps=0.15$ and $\eps=0.11$, respectively.}
    \label{fig:rm_z}
\end{figure}


For another two parameters, namely $r_{\rm M}$ and $z$, their values could not be obtained from low-energy experiments. After sampling the $3 \times 2$ matrix $\mathbf{Y}^\nu$ and selecting out those with two-$\eps$ texture, we show the statistics of $r_{\rm M}$ and $z$ in Fig.~\ref{fig:rm_z}. One obvious feature is that the ratio $r_{\rm M}$ is localized around $r_{\rm M} \simeq 1$, and smaller criteria $\eps$ gives more peaked distribution around $r_{\rm M} \simeq 1$ both for NO and IO. This means that to have a two-$\eps$ texture of $\mathbf{Y}^\nu$, two RH neutrino masses should be of the same order~\footnote{This coincides with the conclusion of Ref.~\cite{Qiu:2021jhi}, which studies the PDF of RH neutrino masses from oscillation data using a different methodology.}.
As shown in Fig.~\ref{fig:rm_z}, large $|z|$ is favored for two-$\eps$ texture and the phase ${\rm Arg}(z)$ has no particular preference, and for a smaller $\eps$ does not bring further obvious trend~\footnote{The small peak in PDF of $|z|$ in IO case at $|z|\sim 0.05$ goes away for smaller $\eps$. So $|z|$ prefers large value for two-$\eps$ texture of $\mathbf{Y}^\nu$.}.
Besides, the distribution of $|z|$ and ${\rm Arg}(z)$ are almost flat in a large region. 
With the above distribution of parameters, one can check the model's compatibility with successful leptogenesis. In the minimum seesaw model, the leptogenesis proceeds via the out-of-equilibrium decays of the heavy neutrinos $\nu_R^1$ and $\nu_R^2$ in the early Universe (in our model $M_1$ tends to equal to $M_2$, see Fig. 3). The generated
lepton asymmetry in such decays is partially converted
into a baryon asymmetry by $(B + L)$-violating sphaleron
processes. The accurate calculation of the leptogenesis with two right-handed neutrinos can be found in Refs. \cite{Barreiros:2018ndn,Antusch:2011nz}, or by using numerical packages such as ULYSSES \cite{Granelli:2023vcm}. By considering only the contribution of $\nu_R^1$, one can simply get a rough estimate of baryon asymmetry, which reads as \cite{Buchmuller:2004nz}
\begin{equation}
\eta_B = \frac{n_B - n_{\Bar{B}}}{n_\gamma} \simeq 10^{-2} \kappa_f \varepsilon_1,
\end{equation}
where $n_B$, $n_{\Bar{B}}$, and $n_\gamma$ are the number density of baryon, antibaryon, and photon, respectively; the efficiency factor $\kappa_f$ represents the effect of washout processes in the plasma and is obtained by solving the Boltzmann equations;  the CP asymmetry parameter $\varepsilon_1$ (only consider $\nu_R^1$'s contribution) is \cite{Buchmuller:2005eh}
\begin{equation}
    \varepsilon_1  = -\frac{3}{16\pi} \frac{M_1}{\left({\Ynu}^T {\Ynu}^*\right)_{11} v^2} \text{Im} \left({\Ynu}^\dagger {\bf d}_{\rm \nu} {\Ynu}^T\right)_{11}.
\end{equation}
Because of the broad distribution of $|z|$ and ${\rm Arg}(z)$ (contained in $\Ynu$), we expect our model will also predict a large range of $\eta_B$. As a cross-check, we use the numerical package ULYSSES to calculate the $\eta_B$, and we find that both NO and IO cases give a wide range of $\eta_B$, i.e.,  $\eta_B \in [10^{-12},  10^{-7}]$, and both can explain the experimental observations, i.e. $\eta_B \simeq 6.2 \times 10^{-10}$ \cite{Komatsu_2011}. As a result, although the parameter distributions of $\delta$, $\alpha$, and $r_M$ are concentrated, the wide distribution of $|z|$ and ${\rm Arg}(z)$ weakens the predictability of the matter-antimatter asymmetry through the leptogenesis mechanism.

\begin{figure}
    \centering
    \includegraphics[width=7cm]{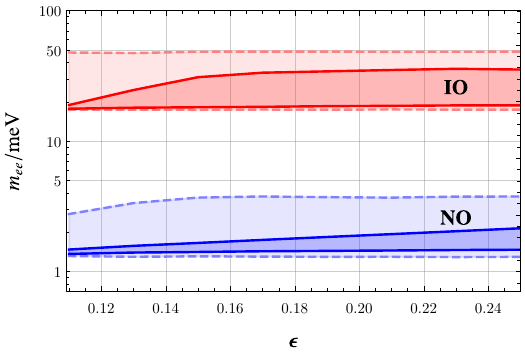}
    \caption{Statistics on the $0\nu\beta\beta$ decay parameter $m_{ee}$ that under the two-$\eps$ texture. The blue band is for NO and the red one is for IO. The darker region is for 1$\sigma$  and the lighter region is 3$\sigma$.}
    \label{fig:mee}
\end{figure}

Under the two-$\eps$ texture, we show the statistics on the parameter $m_{ee} = |\sum_{i} \left(\mathbf{U}_{\rm P}^*\right)^2_{1i} m_i^\nu|$ in Fig.~\ref{fig:mee}, which is crucial to the $0\nu\beta\beta$ experiments.
In the minimal seesaw model, the lightest visible neutrino is massless. So the $m_{ee}$ is nonzero in the NO and IO cases~\cite{ParticleDataGroup:2022pth}.
In our model, $m_{ee}$ tends to become smaller as the criteria $\eps$ both in NO and IO cases. The 1$\sigma$ ranges also become smaller. This is because our $\alpha$ and $\delta$ get more localized PDF as $\eps$ goes smaller as indicated in Fig.~\ref{fig:delta_alpha}.

\begin{figure}[t!]
    \centering
    \includegraphics[width=7cm]{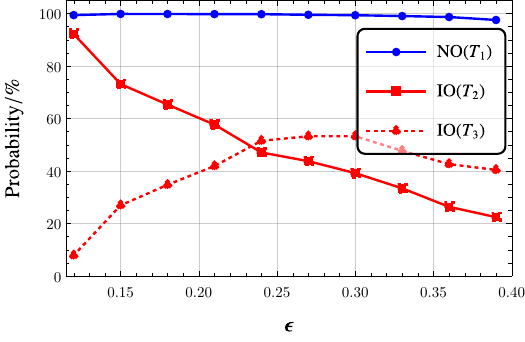}
    \caption{The probability of various textures as a function of $\eps$. The blue-solid line is the $T_1$ texture of the NO case. The red-solid and red-dashed lines are $T_2$ and $T_3$ texture of the IO case, respectively. Other textures merely appear for small $\eps$ and are neglected here. 
    }
    \label{fig:3_texture}
\end{figure}
\begin{figure}
    \includegraphics[width=7cm]{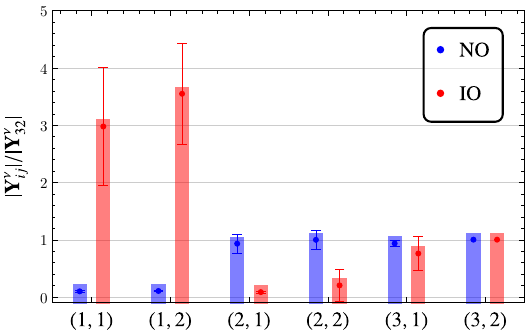}
    \caption{Statistics on the relative norm between the Yukawa matrix $\mathbf{Y}^\nu$ elements with $\eps=0.13$. All elements are rescaled relative to the $|\mathbf{Y}^\nu_{32}|$. Error bars are one standard deviation. Blue labels the NO case and red is the IO case.
    }
    \label{fig:mean_Y}
\end{figure}
The final interesting feature is the specific pattern of two-$\eps$ texture. This is different from the two-zero texture, where the two zero elements can not be located in the same row or column. For two-$\eps$ texture, there is no such constraint. In principle, there are $15$ patterns of two-$\eps$ texture. However, according to our MC simulations, we find that three of them are predominant, they are 
\begin{equation}
    T_1=\begin{pmatrix}
        \eps & \eps \\
        \times & \times \\
        \times & \times
    \end{pmatrix},\;
    T_2 =\begin{pmatrix}
        \times & \times \\
        \eps & \eps \\
        \times & \times
    \end{pmatrix},\;
    T_3 =\begin{pmatrix}
        \times & \times \\
        \eps & \times \\
        \eps & \times
    \end{pmatrix},
    \label{eq:3_texture}
\end{equation}
where `$\times$' stands for parameters whose absolute values are much larger compared to $\eps$s. 
From Fig.~\ref{fig:3_texture}, one could see that $T_1$ texture is the most common texture for the NO case, while $T_2$ and $T_3$ dominate the IO case. Other textures merely appear in the MC simulation. For the NO case, it shows that $T_1$ is always dominant at near $100\%$ and almost independent of $\eps$, while for the IO case, the probability of $T_2$ and $T_3$ is $\eps$-dependent.  At the limit $\eps\to 0$, $T_2$ would be the most common texture. Note that only $T_1$ seems consistent with the hierarchy in standard model quark and charged lepton sectors.

In order to more clearly show the hierarchical relationship between the elements in matrix $\mathbf{Y}^\nu$, we investigate the relative absolute value of each element respective to the $(3,2)$-th element of the Yukawa matrix, i.e., $|\mathbf{Y}^\nu_{ij}|/|\mathbf{Y}^\nu_{32}|$. As shown in Fig.~\ref{fig:mean_Y}, in the NO case, the mean value of the first row elements are of the same order and much smaller than the resting four, which are of the same order as expected. Meanwhile, in the IO case, the first row elements are much larger than the others, while the second row is smallest and the mean value of $|\mathbf{Y}^\nu_{31}|$ is slightly smaller than that of $|\mathbf{Y}^\nu_{32}|$. This agrees with the $T_2$ and $T_3$ pattern as shown in Fig.~\ref{fig:3_texture}. Therefore, our criteria Eq.~\eqref{eq:criteria} successfully select out the desired two-$\eps$ texture.\\

\section{Summary and Discussion} \label{sec:summary}
In this paper, we stick to the minimum seesaw model, where only two heavy right-handed neutrinos are present. This model is quite interesting because it is the simplest model that explains both neutrino masses and matter-antimatter asymmetry problems.
Instead of taking the two-zero texture as previous pieces of literature, such as Refs.~\cite{Harigaya:2012bw,Barreiros:2018ndn}, which are actually not natural from both theoretical and experimental points of view. Here we propose two-$\eps$ texture of the Yukawa matrix. Combining this constraint as well as the neutrino oscillation experiment data, we find that there would be a large CP violation in the NO case and a small CP violation in the IO case. The difference between NO and IO becomes larger when the $\eps$s gets smaller and smaller. Therefore, our model could be well-tested in near-future experiments, such as JUNO and Hyper-Kamiokande. Besides, due to the one vanishing neutrino mass within the minimum seesaw model, total neutrino mass $\sum m^\nu_i$ is almost fixed around 58 meV for NO and 99 meV for IO, which can be tested at cosmological measurements \cite{Lesgourgues:2006nd,TopicalConvenersKNAbazajianJECarlstromATLee:2013bxd,Dvorkin:2019jgs}. 
In addition, the predictions on values of effective Majorana mass $m_{ee}$ are also quite precise, specifically for $\eps=0.11$, we have $m_{ee}\simeq 1.36$--$1.47$ meV for NO and $m_{ee}\simeq 17.69$--$18.89$ meV for IO,
which can be tested at ongoing and/or near-future experiments, such as KamLAND-Zen \cite{KamLAND-Zen:2022tow}, LEGEND-200 \cite{GERDA:2020xhi}, and LEGEND-1000 \cite{LEGEND:2021bnm}.

For the other two parameters, i.e., $r_{\rm M}$ and $z$, our simulations show that two-$\eps$ texture prefers $r_{\rm M}\sim 1$, which means the two right-handed neutrinos have roughly the same mass.
This degenerate spectrum could potentially lower the leptogenesis temperature since the baryon-number asymmetry is $\eta_B \propto (1-r_{\rm M})^{-1}$~\cite{Pilaftsis:2003gt}.
Since the value of Dirac and Majorana phase $\delta$  and $\alpha$ are also approximately fixed, we expected leptogenesis predictions could distinguish two-$\eps$ textures. However, this possibility is lost due to the broad distribution of $z$ (see Fig.~\ref{fig:rm_z}). We have calculated the baryon-to-photon ratio $\eta_B$ for both NO and IO cases with two-$\eps$ texture, we found that both cases can explain observed $\eta_B$. If further constraints can be put on $z$, our model could be distinguished through the leptogenesis mechanism, and we leave this for our future work.\\

\begin{acknowledgements}
T. T. Y. is supported in part by the China Grant for Talent Scientific Start-Up Project and by the Natural Science Foundation of China (NSFC) under Grant No. 12175134 as well as by World Premier International Research Center Initiative (WPI Initiative), MEXT, Japan.

\end{acknowledgements} 


\bibliography{reference}
\bibliographystyle{apsrev}

\end{document}